\documentclass{aa}  

\usepackage{natbib}
\usepackage{graphicx}
\usepackage{txfonts}
\usepackage{hyperref}
\usepackage{mathtools}
\usepackage{subcaption}
\usepackage{balance}

\begin{document} 

  \title{Detectability of shape deformation in short-period exoplanets}
  \author{B. Akinsanmi 
          \inst{1,2,7,\thanks{LSSTC Data Science Fellow}}, S. C. C. Barros\inst{1}, N. C. Santos\inst{1,2}, A. C. M. Correia\inst{3,4,5}, P. F. L. Maxted\inst{6}, G. Bou\'{e}\inst{5} and J. Laskar\inst{5}       
          }

  \institute{  Instituto de Astrofísica e Ciências do Espaço, Universidade do Porto, CAUP, Rua das Estrelas, 4150-762 Porto, Portugal\\
  \email{tunde.akinsanmi@astro.up.pt}
 \and
 Departamento de Física e Astronomia, Faculdade de Ciências, Universidade do Porto, Rua do Campo Alegre, 4169-007 Porto, Portugal
 \and Department of Physics, University of Coimbra, 3004-516 Coimbra, Portugal
\and CIDMA, Department of Physics, University of Aveiro, 3810-193 Aveiro, Portugal
\and ASD, IMCCE, Observatoire de Paris, PSL Universit\'e, Sorbonne Universit\'e, 77 av. Denfert-Rochereau, 75014 Paris, France 
 \and Astrophysics Group, Keele University, Staffordshire ST5 5BG, UK
 \and National Space Research and Development Agency. Airport Road, Abuja, Nigeria.
 }

  \date{Received September 10, 2018; accepted December 06, 2018}

  \abstract
  {Short-period planets suffer from extreme tidal forces from their parent stars. These forces deform the planets causing them to attain non-spherical shapes. The non-spherical shapes, modeled here as triaxial ellipsoids, can have an impact on the observed transit light-curves and the parameters derived for these planets. }
   {We investigate the detectability of tidal deformation in short-period planets from their transit light curves and the instrumental precision needed. We also aim to show how detecting planet deformation allows us to obtain an observational estimate of the second fluid Love number from the light curve which gives valuable information about the planet's internal structure.}
   {We adopted a model to calculate the shape of a planet due to the external potentials acting on it and used this model to modify the \textit{ellc} transit tool. We used the modified \textit{ellc} to generate the transit light curve for a deformed planet. Our model is parameterized by the Love number, hence for a given light curve we can derive the value of the Love number that best matches the observations.}
   { We simulated the known cases of WASP-103b and WASP-121b which are expected to be highly deformed. Our analyses showed that instrumental precision $\leq$\,50\,ppm/min is required to reliably estimate the Love number and detect tidal deformation. This precision can be achieved for WASP-103b in $\sim$40 transits using the Hubble Space Telescope and in $\sim$300 transits using the forthcoming CHEOPS instrument. However, fewer transits will be required for short-period planets that may be found around bright stars in the TESS and PLATO survey missions. The unprecedented precisions expected from PLATO and JWST can permit the detection of shape deformation with a single transit observation. However, the effects of instrumental and astrophysical noise must be well-considered as they can increase the number of transits required to reach the 50\,ppm/min detection limit. We also show that improper modeling of limb darkening can act to bury signals related to shape of the planet thereby leading us to infer sphericity for a deformed planet. Therefore accurate determination of the limb darkening coefficients is required to confirm planet deformation.}
   {}

\titlerunning{Detectability of shape deformation}
\authorrunning{B. Akinsanmi et al.}
\maketitle


\section{Introduction}
The existence of planets with short-period orbits around their stars came as a surprise at the inception of exoplanet discoveries especially because the first case was a gas giant \citep{mayor} bearing no resemblance to the planet configuration in our Solar System. Several of these planets have now been found as they represent some of the most easily detected planets using both the transit and radial velocity methods. Planets reach their final shapes having attained hydrostatic equilibrium from balancing gravitational, pressure and other external forces acting on them. Planet shapes are often assumed to be spherical for simplicity but they are triaxial in reality. For very short-period planets (P < 1-2 days), the close proximity to their stars exposes them to strong tidal forces which deforms them and increases the triaxiality of their equilibrium shapes. Contribution to deformation can also come from planet's rotation which makes the planet oblate \citep{barnes03}.

Planet shape can have noticeable effects on the light curve obtained from transit observations \citep{seagerhui,carterwinnb,carterwinna}. Analysis of transit light curve of planets assuming planet sphericity allows for a spherical radius $R_{spr}$ to be obtained. However, \citet{leconte} showed that planet deformation due to tidal and rotational forces lowers the observed transit depth in comparison to a spherical planet. This causes an underestimation of the planet's radius when sphericity is assumed in the transit light curve analysis of a deformed planet. Since the planet's density is calculated from the assumed spherical radius, the obtained density will consequently be overestimated. \citet{burton} thus provided density corrections for some short-period planets expected to be tidally deformed based on the Roche approximation \citep{chand}. Tidal deformation is particularly significant for planets orbiting close to their stellar Roche limits and a number of planets have been discovered to orbit so close to this limit that they are at edge of tidal disruption (e.g. \citealt{ gillon14,delrez}). For some of these planets, theoretical calculations have been done using the Roche model by \citet{budaj} to estimate the planet shape and correct the derived spherical radii and densities for the expected planet deformation (e.g. \citealt{ southworth15,delrez}). \citet{correia14} formulated an analytical model for computing the shape of a deformed planet based on the fluid second Love number and also showed the difference between light curves of deformed and spherical planets.

Despite these efforts, there has been no observational detection of tidal deformation in short-period planets which would provide better estimates of their parameters.  We therefore investigate the possibility of detecting deformation in the transit light curve of short-period planets with some current and near-future observational instruments. We modify the \textit{ellc} transit tool by \citet{maxted} \footnote{Available at \href{https://pypi.org/project/ellc/}{ https://pypi.org/project/ellc/}} to incorporate the planet shape model by \citet{correia14}. The modified \textit{ellc} is used to generate the light curve for a deformed planet based on its fluid second Love number. This allows us to obtain an estimate for the planet's Love number that best matches the transit observations which provides insights on the internal structure differentiation of the planet.

In Sect. 2, we summarize the model used to compute the shape of the planet and modification of the transit tool used to generate the light curves. In Sect. 3 we apply the modified tool to investigate the detectability of planet deformation taking case study of a known short-period planet. In Sect. 4, we discuss the results and some useful considerations for detecting planet deformation. We present our conclusions in the last section.

\begin{figure}
\centering
\includegraphics[width=1\linewidth]{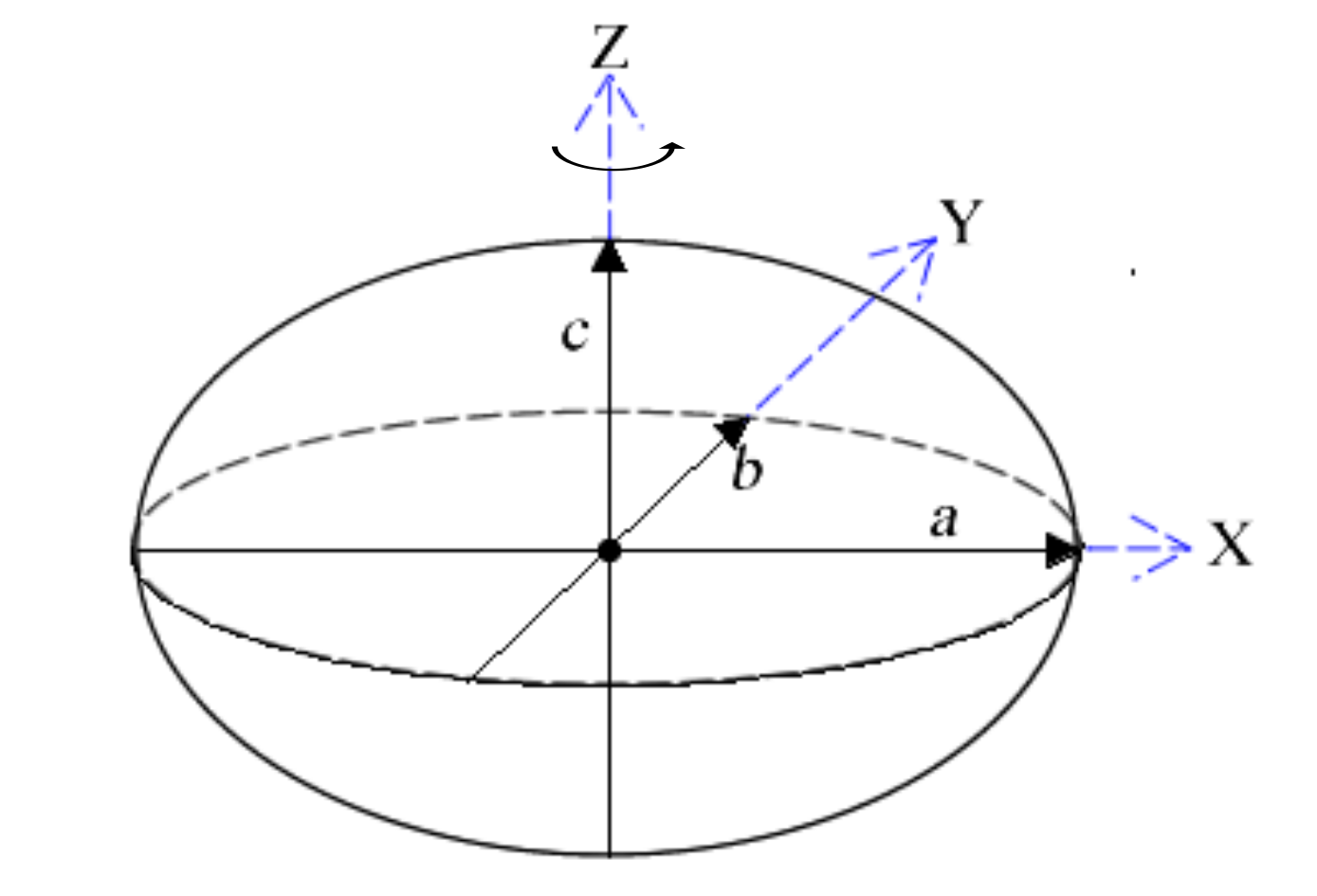}
\caption{Schematic of triaxial ellipsoid centered on the origin of the Cartesian coordinate system ($X, Y, Z$) with positive X-axis pointing towards the star.}
\label{coord}
\end{figure}

\section{Modeling transit of deformed planets}
\subsection{Planet shape}
Modeling the shape of a deformed planet follows the analytical formulation by \citet{correia14} in which the planet is described by a triaxial ellipsoid centered at the origin of a Cartesian coordinate. As shown in Fig. \ref{coord}, the semi-principal axes ($a, b, c$) of the ellipsoid are aligned with the $X, Y, Z$ axes of the coordinate system respectively. The equilibrium shape and mass distribution of a planet depends on the forces acting on it which are the planet's self gravity and other perturbing potentials. The planet can deform under the influence of centrifugal and tidal potentials. For a tidally locked close-in planet with circularized orbit of radius $r_{0}$, \citet{corr13} give the non-spherical contribution of the perturbing potential on the planet's surface as

\begin{equation}
    V_{p}=\frac{1}{2}\Omega^{2}Z^2-\frac{3GM_{\ast}}{2r_{0}^3}X^2,
\end{equation}

\noindent where G is the gravitational constant. The first term on the right-hand-side is the deformation contribution from the centrifugal potential resulting from planet's coplanar  and synchronous rotation rate $\Omega$ about the Z-axis. The second term refers to the tidal contribution to the deformation along the X-axis by a star of mass $M_{\ast}$.

Following \citet{love}, \citet{correia14} describes this deformation using a Love number approach such that the fluid second Love number for radial displacement $h_{f}$ is related to the radial deformation of the planet $\Delta R$. The equilibrium surface deformation is thus given by 
\begin{equation}
    \Delta R= -h_{f}V_{p}/g,
\end{equation}

\noindent where $g$ is the average surface gravity of the planet. $h_{f}$ is a dimensionless quantity that quantifies a planet's response (deformation) to a perturbing potential\footnote{$h_{f}$=1+$k_{f}$ where $k_{f}$ is the fluid second Love number for potential \citep{corr_boue}. Calculation of the different Love numbers can be found in \citet{sabadini}.}. The magnitude of $h_{f}$ depends on the mass distribution of the planet. More homogeneous planets have higher $h_{f}$ whereas planets that are more centrally condensed have lower $h_{f}$ \citep{kramm11,kramm12}. For an incompressible homogeneous planet, $h_{f}$ = 2.5 which is the theoretical maximum value \citep{leconte,correia14}. The physical values of $h_{f}$ range from 1 to 2.5 where $h_{f}=1$ would represent highly differentiated bodies with high core mass like FGK stars and $h_{f}=2.5$ is only possible for significantly homogeneous bodies like asteroids.  In comparison Jupiter has $h_{f}\approx1.5$ and Earth has $h_{f}\approx2$ \citep{yoder}. First observational measurement of Saturn's Love number was recently obtained by \citet{LAINEY2017} leading to a value of $h_{f}=1.39$ (from $k_{f}=0.39$).

Due to the synchronous rotation, the semi-principal axis $a$ of the planet always points in the direction of the star leading to a tidal deformation along $a$. 	The shape of the planet is such that $a>b>c$ and the deformation is kept constant along the circularized orbit. For the ellipsoid, we can define also the radius of a sphere that will enclose the same volume as the ellipsoid so that $R_{v}=(abc)^{1/3}$. According to the formulation by \citet{correia14}, the semi-principal axes are related as $a=b\,(1+3q)$ and $c=b\,(1-q)$. We can then write $b$ as a function of $R_{v}$ to first order in the parameter $q$ as
\begin{equation}
\label{b}
    b\simeq R_{v}\,\left( 1-\frac{2}{3}q\right)\,, 
\end{equation}
so that
\begin{equation}
\label{a}
    a=b\,(1+3q)\simeq R_{v}\,\left( 1+\frac{7}{3}q\right)  
\end{equation}
and 
\begin{equation}
\label{c}
    c=b\,(1-q) \simeq R_{v}\,\left( 1-\frac{5}{3}q\right), 
\end{equation}

\noindent where $q$ is an asymmetry parameter that relates to $h_f$ according to
\begin{equation}
\label{qq}
    q=\frac{h_{f}}{2}\frac{M_{\ast}}{m_{p}}\left(\frac{R_{v}}{r_{0}}\right)^3.
\end{equation}

\begin{figure}[t!]
\centering
\includegraphics[width=1\linewidth]{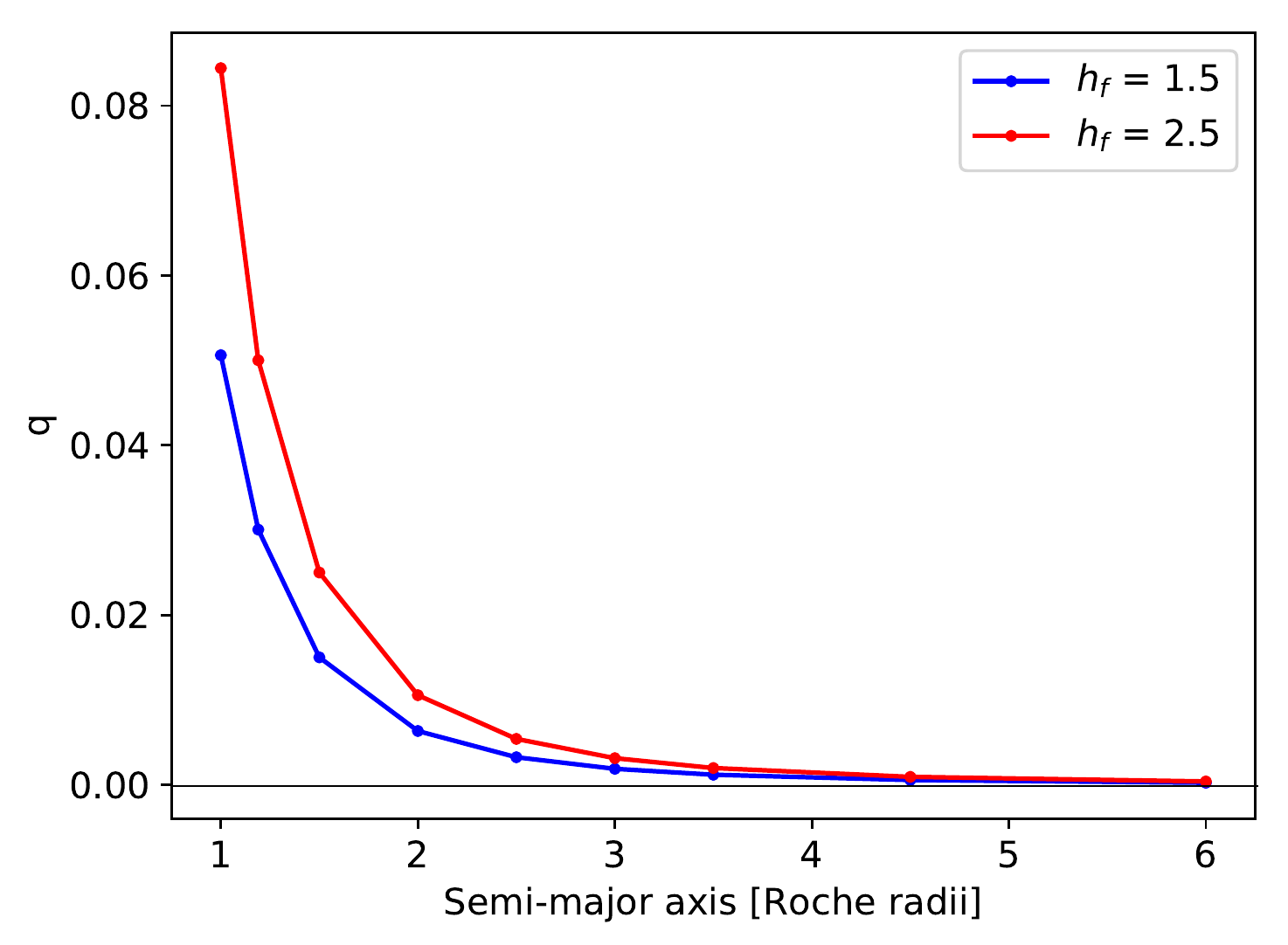}
\caption{Quantification of tidal deformation as a function of distance to the star for two different $h_{f}$ values.}
\label{q_comp2}
\end{figure}

The asymmetry parameter $q$ quantifies the deformation of a planet, i.e., the difference between the ellipsoid's semi-principal axes. Maximum deformation (hence maximum $q$) is attained for a given planet when it orbits at the Roche radius ($r_{0}=r_{R}=2.46\,R_{v}[M_{\ast}/m_{p}]^{1/3}$). Therefore, for maximum $h_{f}$ = 2.5, we have $q_{max} \simeq 0.083$. The equilibrium shape of a planet thus depends on its radius, the planet's fluid second Love number $h_{f}$, the mass ratio between star and planet $M_{\ast}/m_{p}$ and also the planet distance $r_{0}$ from the star. Figure \ref{q_comp2} shows how tidal deformation becomes negligible with semi-major axis (in units of its Roche radii) for a given body with $h_{f} = 2.5$ and again with Jupiter's $h_{f} = 1.5$. We see that far away from the star, irrespective of the value of $h_{f}$, the planet does not deform ($q\simeq0$) and so its shape remains largely spherical ($a\simeq b\simeq c$ from Eqs. \ref{b}-\ref{c}). In general, Eq. \ref{qq} shows that tidal deformation is more relevant for large planets orbiting very close to their Roche radii. Planets with the highest absolute deformation (highest product $q\times R_{v}$) present the best chances to detect deformation.

\subsection{Transit model}
Planetary features that change the shape of a planet (oblateness or rings) have the effect of modifying the transit light curves (e.g. \citealt{barnes03,akin18}). In the same vein, tidal deformation of a planet can modify the observed transit light curve. To model the transit of a deformed planet, the above ellipsoidal shape model by \citet{correia14} was incorporated as a subroutine into a new version of the \textit{ellc} transit tool by \citet{maxted}. The \textit{ellc} light curve model allows 
the projection of the ellipsoid and generation of the corresponding transit light curve. The projected shape of the ellipsoid on the stellar disk is an ellipse whose dimensions depend on the phase of the planet due to rotation of the ellipsoid with phase (\textit{see} Fig. A.1 in \citealt{correia14}). The rotation of the ellipsoid causes the cross-section of the planet to vary during transit. It should be noted that the shape correction model by \citet{budaj} does not account for the varying ellipsoidal cross-section during transit thereby making \textit{ellc} a more complete model involving this observational effect. Detailed descriptions of the \textit{ellc} tool and the input parameters can be found in \citet{maxted}.

The modified transit model, in addition to usual transit parameters, takes the value of $h_{f}$ and the ellipsoid's volumetric radius $R_{v}$ as inputs. Therefore, by fitting the ellipsoidal model to the transit observation, all the parameters of the transit, including the shape of planet, can be obtained and $h_{f}$ is estimated from the best fit of the model. So rather than obtaining the usual transit radius $R_{spr}$ from spherical planet models, we obtain the best-match dimensions $a, b, c$ of the ellipsoidal planet and calculate $R_{v}$.

\begin{figure}[tp!]
\centering
\includegraphics[width=1\linewidth]{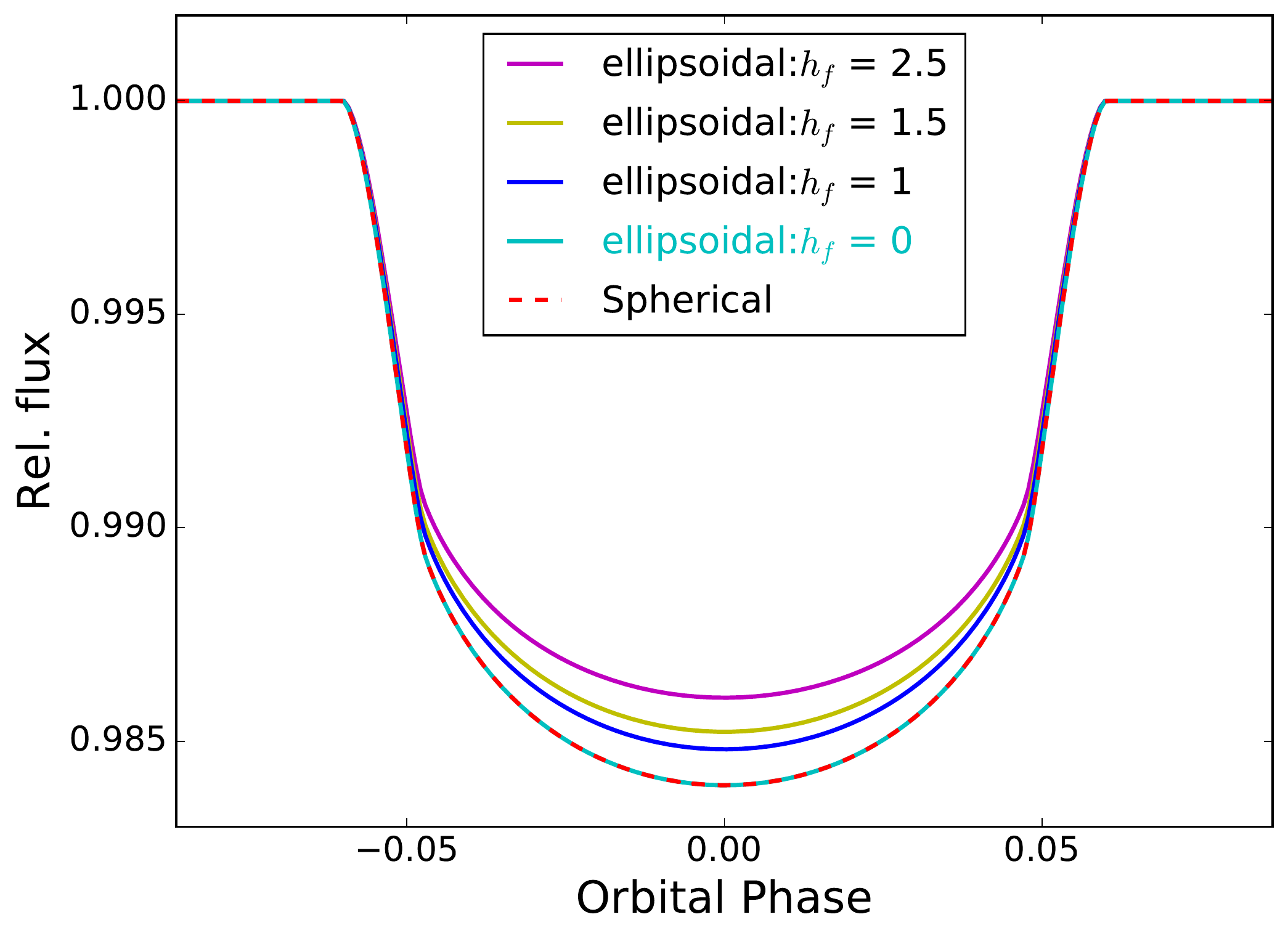}
\caption{Comparison of ellipsoidal model light curves of different $h_{f}$ values with spherical model light curve for WASP-103b.}
\label{hfcomp}
\end{figure} 

\begin{table}[]
\caption{System parameters for WASP-103b.}
\centering
\label{parameters}
\begin{tabular}{lll}
\hline \hline
\multicolumn{1}{l|}{Quantity {[}Unit{]}} & \multicolumn{1}{l|}{Symbol\qquad\qquad} & \multicolumn{1}{l}{Value\qquad\qquad} \\ \hline
Planet radius {[}$R_{\odot}${]}           & $R_{spr}$                     & 0.1604                     \\[3pt]
Planet mass {[}$M_{\odot}${]}             & $m_{p}$                     & 0.0014                     \\[3pt]
Stellar radius {[}$R_{\odot}${]}          & $R_{\ast}$                  & $1.4130$                   \\[3pt]
Stellar mass {[}$M_{\odot}${]}            & $M_{\ast}$                  & $1.2050$                    \\[3pt]
Semi-major axis {[}$R_{\odot}${]}\qquad\qquad         & $r_{0}$                     & 4.2555                     \\[3pt]
Roche radius {[}$R_{\odot}${]}            & $r_{R}$                     & 3.7534                    
\\ \hline
\end{tabular}
\end{table}

\subsection{The case of WASP-103b}
\begin{figure}[tp!]
\centering
\includegraphics[width=1\linewidth]{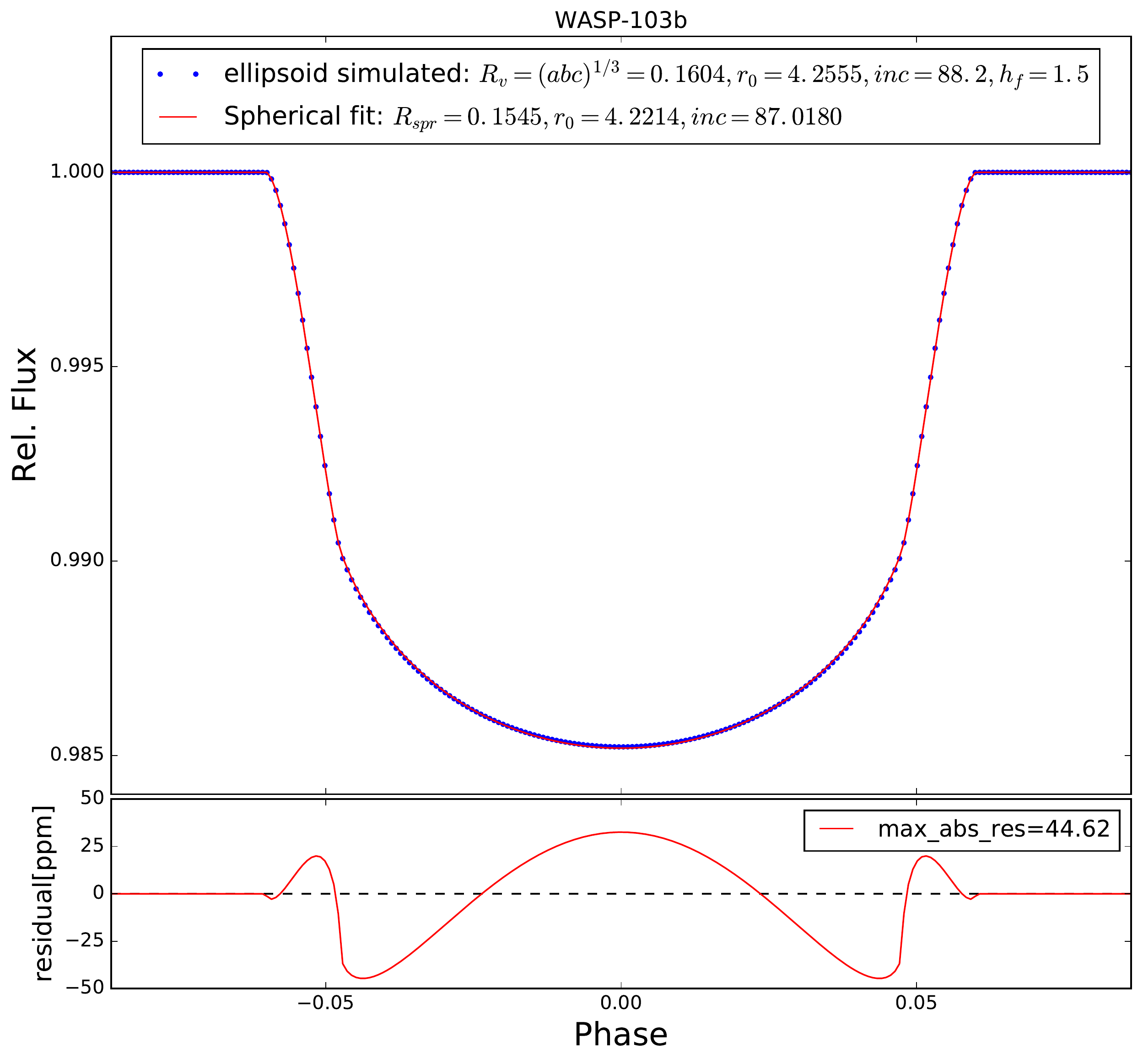}
\caption{Spherical fit to simulated deformed WASP-103b light curve. The bottom plot shows the residual representing the signature of deformation with amplitude quoted as the maximum absolute residual (max\_abs\_res). All length measurements are given in units of Solar radii.} 
\label{sphrfit}
\end{figure}

To illustrate the output of \textit{ellc} for an ellipsoidal planet, we take the case of WASP-103b, an ultra-short-period planet (P=22.2\,hr) reported to be on the edge of tidal disruption \citep{gillon14} making it an ideal candidate to detect deformation. Based on revised parameters by \citet{southworth16}, it has an average radius of 1.596\,$R_{Jup}$ and mass of 1.47\,$M_{Jup}$ (Table \ref{parameters}). It orbits its star at a semi-major axis ($r_{0}$) of 0.01979\,AU and an inclination ($inc$) of $88.2^{\mathrm{o}}$. It is assumed to be on the edge of tidal disruption due to its semi-major axis of only 1.13 times its Roche radius. Taking the quoted radius as the volumetric radius of the ellipsoid, Fig. \ref{hfcomp} compares the spherical planet light curve for WASP-103b to its ellipsoidal counterparts with different $h_f$ values. It is seen that the light-curve of the ellipsoidal model changes noticeably for different values of $h_{f}$ and also compared to the spherical case. This is because the ellipsoidal planet projects only a small cross-section of its shape during the transit thereby leading to a lower transit depth when compared to the spherical planet. The mid-transit phase has the smallest ellipsoidal cross-section of $bc\simeq R_{v}^{2}(1-7q/3)$ which is less than the cross-section $R_{spr}^2$ if the planet were spherical. Therefore, if a spherical model is used to make a fit to the observation of an ellipsoidal planet, the spherical radius $R_{spr}$ derived will be smaller than the actual volumetric radius $R_{v}=(abc)^{1/3}$ of the ellipsoid (see Fig. \ref{sphrfit}). This is in agreement with the result from \citet{leconte}. Differences in transit depth as $h_{f}$ varies in Fig. \ref{hfcomp} is because higher $h_{f}$ for the same planet causes more deformation which leads to even smaller projected cross-sectional area. In our code, we allow for a case where $h_{f}$\,=\,0 (although not physical) to imply no deformation for the planet so that the ellipsoidal planet model is equivalent to that of a spherical planet and they produce the same light-curve with $R_{v}=R_{spr}$. This is important for the analysis we perform in the next section and allows us to use the same model to explain both a deformed and a spherical planet. \citet{maxted} already showed that the spherical light curve of \textit{ellc} is in agreement with other transit tools like \textit{BATMAN} \citep{kreid}.

\subsection{Signature of deformation in transit light curves}
\label{2.4}
 
Figure 1 in \citet{correia14} showed difference plots between ellipsoidal and spherical light curves assuming both planets cover the same stellar area at the start of transit (full ingress). This perfectly captures the flux variation induced by deformation as both planets transit but is not the signature one will obtain from real observations since the transit parameters will be initially unknown and must be determined from a fitting process. The observable signature of planet deformation is the residual between the deformed planet's light curve and the best-fit spherical model. In Fig. \ref{sphrfit}, we simulated the light-curve of deformed WASP-103b using our ellipsoidal model with parameters given in Table \ref{parameters} and performed least-squares fitting using a spherical planet model. The residual from the fit is shown in the bottom panel and it represents the signature of deformation for the simulated planet.The parameters derived from the fitting process are systematically wrong as they adjust to mimic the signature of deformation. This also shows that the assumption of sphericity for a planet affects not only the radius derived but also the other transit parameters and models that adjust only this radius are incomplete. We see in the residuals that the signature of deformation manifests in two regions. First is at ingress and egress phases owing to oblateness ($b > c$) of the planet as identified in previous studies (e.g. \citealt{seagerhui,barnes03}). A second prominent feature is seen as a bump centered on the mid transit phase due to the varying star eclipsed area caused by ellipsoid rotation as it transits. This second feature is as a result of tidal deformation which was not accounted for in the previous studies mentioned but manisfests in our model due to full projection of the ellipsoidal shape as it rotates with phase \citep{correia14}.

\begin{figure}[tp!]
\centering
\includegraphics[width=1\linewidth]{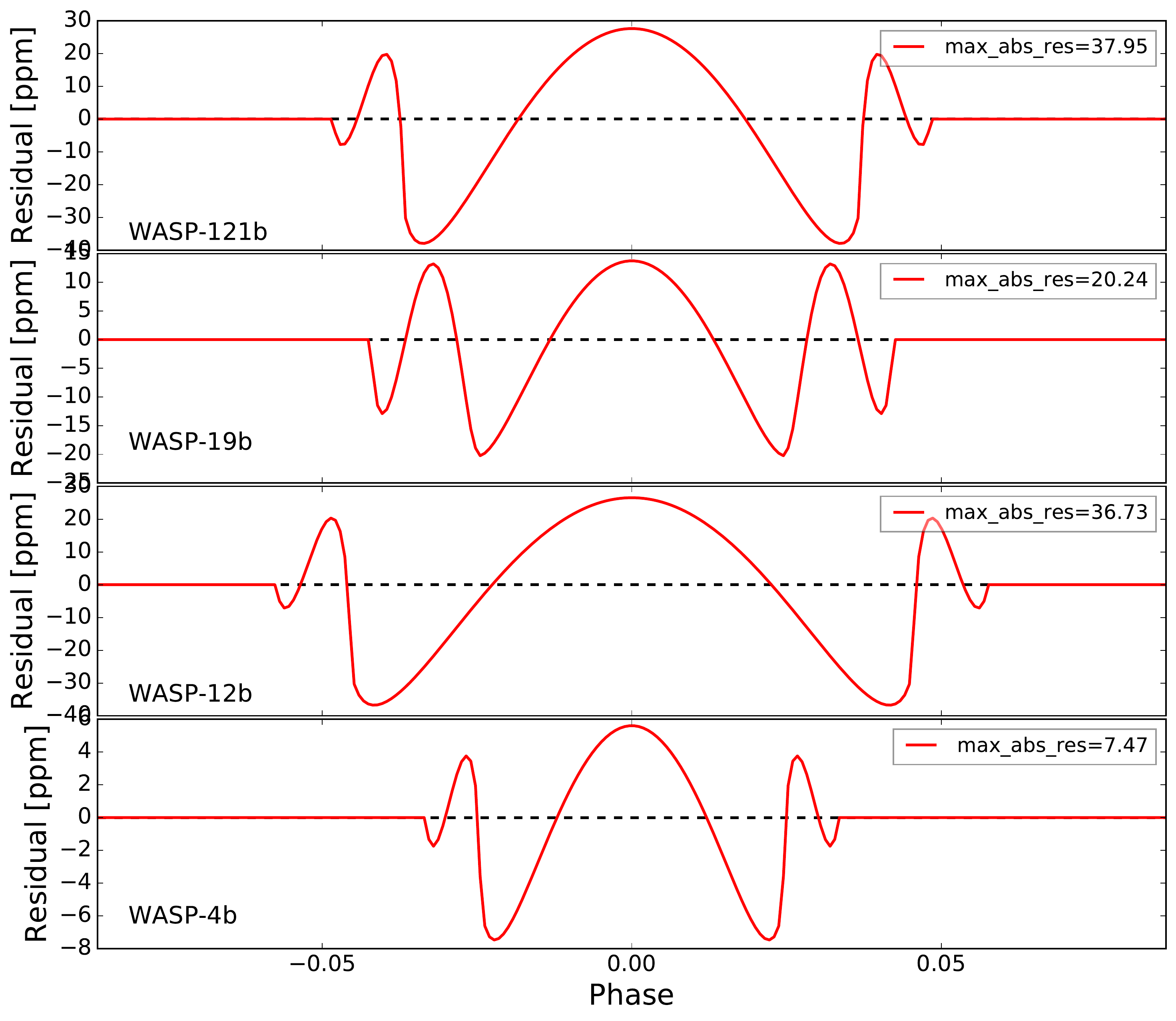}
\caption{Residuals from spherical fit to ellipsoidal simulations of different short-period planets in comparison to WASP-19b, 12b and 4b from \citet{correia14}}
\label{diff_fit}
\end{figure}

To compare the deformation signal obtained from the fitting process with the flux difference plot in \citet{correia14}, we perform spherical fits to the ellipsoidal simulation of other short-period giant planets WASP-19b, WASP-12b, WASP-4b and WASP-121b that were presented in the study and expected to be deformed. The residuals are shown in Fig. \ref{diff_fit}. We see from Figures \ref{sphrfit} and \ref{diff_fit} that the amplitude of the deformation signature is just about 40\,ppm for the most deformed planets (WASP-103b and WASP-121b) while the amplitude from the difference curves in \citet{correia14} are up to 100\,ppm. We reiterate that the latter should not be taken to imply high signal detectability. 

WASP-103b, WASP-121b and WASP-12b have the highest residual amplitudes and therefore present the best possibility of detecting deformation. Other planets likely to be deformed are HATS-18b, WASP-76b and WASP-33b but have lower residual amplitudes of 20, 14 and 12\,ppm respectively.

\section{Detectability of planet deformation and measurement of planet Love number}
 
The residuals of the spherical fit to a deformed planet's light curve is informative in detecting deformation as it shows that the spherical model does not fully explain the observation. However, some of the signature of the deformation is masked in the errors of the parameters obtained. To correctly estimate the planet transit parameters, our ellipsoidal model can be used to fit the transit observation. In doing so, we also obtain a value for the Love number that best fits the observation if there is enough precision in the data. The benefit of this approach is that we can fit the ellipsoidal model to any transit observation and, by the value of $h_{f}$ recovered, ascertain if planet deformation is detectable or not. If we cannot detect the deformation, we get $h_{f} \approx 0$ which as shown in Fig. \ref{hfcomp} is equivalent to the fit of a spherical planet model.

\begin{figure}[t!]
\centering
\includegraphics[width=1\linewidth]{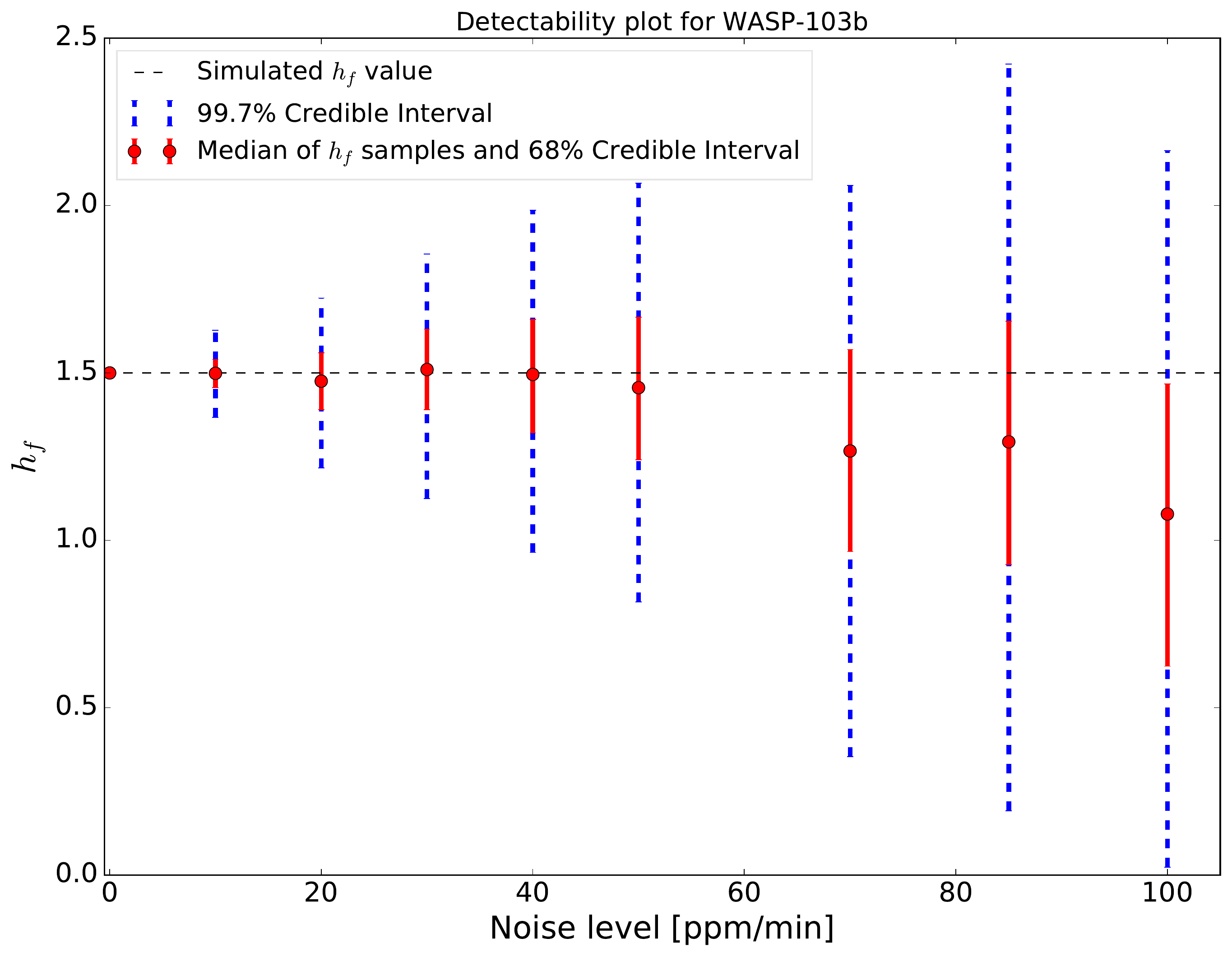}
\caption{Detectability of deformation in WASP-103b considering different noise levels. The black dashed line is the simulated $h_{f}$ value. The points are the median of the $h_{f}$ samples at each noise level. The red error bars show the 68\% credible interval ($\simeq \pm1\sigma$) while the blue error bars show the 99.7\% credible interval ($\simeq \pm3\sigma$). }
\label{detect}
\end{figure}

Therefore, detectability of tidal deformation using the ellipsoidal model relies on the ability to recover a non-zero value of $h_{f}$ with statistical significance from a fitting process. Despite being able to infer deformation with only detection of $h_{f} \gg 0$, we will need to have $h_{f} \geq 1$ with some significance where the values give actual physical interpretation to astronomical bodies. To illustrate the detectability, we created simulated observations of deformed WASP-103b with 1\,min cadence using its parameters as stated above with $h_{f}$\,=\,1.5. We used the Limb darkening toolkit (\textit{ldtk}) by \citet{pav15} to compute quadratic limb darkening coefficients of [0.5343, 0.1299] and their uncertainties [0.0012, 0.0027] in the CHEOPS bandpass for the star with stellar parameters given in \citet{gillon14}. We added random gaussian noise of different levels to the simulated data in each test run. We then investigated how well we can recover the value of $h_{f}$ and at what noise level it would be impossible to distinguish between the light curve of a spherical planet and that of a deformed planet. This is important to know the instrumental precision necessary to detect deformation in close-in planets.

We performed Markov Chain Monte Carlo (MCMC) analysis to estimate the transit parameters and their uncertainties using the \textit{emcee} package \citep{foreman} with uniform priors on $h_{f}$ in the range [0, 2.5]. As shown in Appendix \ref{corn}, when a noise level of 30\,ppm is added to the simulated observation, $h_{f}$ is reliably recovered with 99.7\% of its samples (within $\simeq \pm3\sigma$) greater than 1. This proves that the result is statistically significant and implies that the planet is indeed deformed. Moreover, the residual from the fit does not show any structure related to the deformation signal. However, when a noise level of 100\,ppm is added to the observation the median of the distribution suggests a deformed planet but because its width encompasses $h_{f} = 0$ (spherical model), planet deformation cannot be asserted (Appendix \ref{corn100}). Figure \ref{detect} shows the detectability plot summarizing the results for the different noise levels added to the observation. We see that the significance of $h_{f}$ detection above 1 reduces as the noise level of the observation increases. For instance, at 50\,ppm noise level, $h_{f}$ samples are well above 0 implying that the ellipsoidal model provides a better fit than the spherical model. However, the samples with $h_{f}$ < 1 do not represent physical values for a planet but the detection still gives $\sim 95\%$ of the samples above 1. Beyond 50\,ppm, fitting the observation with a spherical model becomes increasingly more probable. With noise levels as high as 100\,ppm, the spherical and ellipsoidal models produce comparable fits.

\begin{table}[]
\centering
\caption{Number of transits required to reach 50ppm/min noise level with CHEOPS and PLATO for different stellar magnitudes.}
\label{noise}
\begin{tabular}{llc|llc}
\hline\hline
\multicolumn{3}{c|}{CHEOPS}                                                                & \multicolumn{3}{c}{PLATO}                                                                                                                   \\
\hline
\multicolumn{1}{c}{$m_{V}$} & \multicolumn{1}{c}{Noise/min} & \multicolumn{1}{c|}{\# transits} & \multicolumn{1}{c}{$m_{V}$} & \multicolumn{1}{c}{Noise/min} & \multicolumn{1}{c}{\begin{tabular}[c]{@{}c@{}}\# transits\end{tabular}} \\
\hline
6.5                     & 150\,ppm       & 9         &  8      &  62\,ppm        &  2                                                                                 \\[5pt]
8                       & 186\,ppm       & 14        &  10     &  209\,ppm       &  17                                                                                \\[5pt]
10                      & 319\,ppm       & 40        &  11	  &  263\,ppm	   &  28  \\[5pt]
12                      & 855\,ppm       & 293       &  13	  &  619\,ppm	   &  153	
\\ \hline
\end{tabular}
\tablefoot{Noise levels of CHEOPS was obtained from CHEOPS science team (private communication) and that for PLATO was converted to ppm/min from  \citealt{rauer} (Table 2 and Fig. 14 ).}
\end{table}


\section{Discussion}

The results show that noise levels below 30\,ppm offer the best chance at detecting deformation for our test case of WASP-103b since we retrieve $h_{f}$  with $\geq3\sigma$ significance above 1. However, we could define a lower limit on our detection confirmation such that we require to have ($h_{f}-1\sigma) \geq 1$ which puts 84\% of the recovered $h_f$ samples in physical values expected for planets. This is satisfied for noise levels of 50\,ppm and below. 

A photometric precision of 50\,ppm/min is not yet attainable using current observational instruments. For our case system, WASP-103 is a $12^{th}$ magnitude star and the photometric precision to be attained by the near-future instrument CHEOPS for this star is 855\,ppm per minute. Attaining a reduced photon noise level of 50\,ppm/min for this star using CHEOPS requires $\sim$293 transit observations of WASP-103b. For the interesting candidate WASP-121b which orbits its star of magnitude $m_{V}$=10 \citep{delrez}, our analysis also showed detectability of deformation with 50\,ppm/min noise level. CHEOPS precision for a $10^{th}$ magnitude star is 319\,ppm/min thereby requiring only 40 transit observations to detect deformation in this planet. Although information from the CHEOPS consortium indicates that WASP-121 might not be in the visibility region, new interesting planet candidates with short period orbits may appear from future surveys targeting bright stars, such as PLATO \citep{rauer} and TESS \citep{ricker}. For these planets around stars brighter than $m_{V} = 9$, we expect photon noise levels as low as 150\,ppm/min with CHEOPS \citep{broeg} and <\,62\,ppm/min with PLATO \citep{rauer} and thus require fewer transits to reach the 50\,ppm limit needed to detect planet deformation as reported in Table \ref{noise}. For these stars, TESS will have a relatively higher noise level of 464\,ppm/min \citep{sullivan} which is not desirable for detecting deformation. Observations with the forthcoming JWST will also be immensely beneficial as it is expected to attain photon-noise floor $\sim$40\,ppm (65\,secs) on its NIRCam instrument amongst others \citep{beichman}. Attainment of this noise level implies that only one transit observation will be required in order to detect tidal deformation in a suitable short-period planet.

Unfortunately, interesting short-period planets expected to be significantly deformed were not found within the original \textit{Kepler} survey field which would have allowed several transit observations of any found target. The WFC3 instrument on the Hubble Space Telescope (HST) achieved noise level of 172\,ppm (103\,secs) for 2 full-orbit observations of WASP-103 \citep{kreid18}. Therefore, with $\sim$40 transits of WASP-103b using HST, we can attain the required precision of 50\,ppm/min.  However, some factors can still affect the detectability of deformation, some of which are mentioned below.\\

\noindent \textit{Temporal Resolution}: The temporal resolution of the observation can affect the detectability. We have used 1\,min cadence in our simulations to enable good resolution of the ingress and egress phases which have short durations especially for these short period planets. A lower cadence than this reduces the precision with which $h_{f}$ and other parameters are recovered. At the 30\,ppm noise level, changing the cadence from 1\,min to 4\,mins and 8\,mins increases the error on $h_{f}$ from $\pm0.12$ to $\pm0.23$ and $\pm0.38$ respectively.\\
 
\noindent \textit{Orbital inclination}: The inclination of the orbit plays a role in the signature of deformation. Lower inclinations indicate a shorter transit duration so the effects referred to in residuals of Fig. \ref{sphrfit} and Sect. \ref{2.4} will be shorter in time making them more difficult to be temporally resolved especially at the ingress and egress phases. In addition, a longer transit duration allows the projected ellipse area to vary more (longer phase rotation of ellipsoid) making the light-curve more markedly different from that of the spherical planet thereby leading to a higher amplitude bump around the mid transit (\textit{see also} Fig. A.1 in \citealt{correia14}). The effects of deformation in light curves is maximum at inclination of $90^{\mathrm{o}}$ where $h_{f}$ is recovered with the best precision.\\

\noindent \textit{Limb darkening coefficients (LDCs)}: As shown in Fig. \ref{sphrfit}, the signature of deformation is prominent at ingress and egress phases with a bump centered around the mid transit phase. The stellar limb darkening affects light curves similarly in these regions (see effects of LDC modeling in \citealt{neilson}), so we tested the impact of inaccurate estimation of limb darkening coefficients on the recovery of $h_f$ from the light curve. This was attempted on the 30\,ppm noise level simulation in two ways and the result summarized in Table \ref{LDCtests}. First we fixed the limb darkening coefficients to wrong values that are slightly different from the true values used to generate the simulated observation. We found that for wrongly fixed LDC values which are smaller than the true values, the signature of deformation gets damped as we recover lower $h_{f}$ values than simulated. When the values are fixed at values up to 0.01 smaller than the truth, the entire $h_{f}$ distribution falls around 0 and we infer a spherical planet (see left plot in Fig. \ref{LDCs}). On the other hand, $h_{f}$ values are amplified when LDCs are fixed at values higher than the truth. For LDC values fixed at 0.015 higher than the truth, the recovered $h_{f}$ distribution is pushed towards the maximum of 2.5. In the later case, we can infer that the planet is deformed but cannot ascertain the extent of deformation due to inaccurate estimation of $h_{f}$ which is evident from the obtained marginalized distribution (see right plot in Fig. \ref{LDCs}). The other attempt was to fit the LDCs by including them in the hyperparameters. We use a gaussian prior with the true LDC values as mean and $\sigma=0.01$. The MCMC sampling produced a wide $h_f$ distribution centered close to the true value but with errors as large as $\pm 0.4$ (left plot in Fig. \ref{LDC_priors}) making it difficult to ascertain planet shape. However, when tighter priors (e.g. using errors obtained from deriving LDCs with \textit{ldtk}) are imposed on the LDCs, $h_{f}$ is well-recovered with errors of just $\pm0.18$ to infer deformation (right plot in Fig. \ref{LDC_priors}). It should be noted that the LDC error estimates from \textit{ldtk} are very small and have often had to be inflated in literature during fitting to account for systematic errors in the atmospheric models (e.g. \citealt{raynard,maxted18b}).

Alternatively, the power-2 limb darkening law has been recommended for the analysis of transit light curves as it has been shown to provide remarkable agreement between stellar atmospheric models and observations, particularly for cool stars \citep{morello,maxted18}. The transformation of the two parameters of the power-2 law in \citep{maxted18} reduces the correlation between them during fitting and small errors of [0.011, 0.045] can be obtained on them. The fitting process can attempt different LDC laws so that the law with the best match to the observation and that produces the least errors on the derived parameters will be preferred. \\ 

\begin{table}[]
\centering
\caption{Results of LDC tests and $h_{f}$ values recovered. The plots are shown in Figures \ref{LDCs} and \ref{LDC_priors}}
\label{LDCtests}
\begin{tabular}{llc}
                \hline\hline
LDC tests            & Values            & $h_{f}$ recovered           \\
                \hline
Fixed at 0.01 below  & {[}0.5243,0.1199{]}   & $0.12^{+0.11}_{-0.08}$ \\[5pt]
Fixed at 0.015 higher & {[}0.5493,0.1449{]}       & $2.44^{+0.04}_{-0.06}$ \\[5pt]
Gaussian priors      & \begin{tabular}[c]{@{}l@{}}Mean={[}0.5343,0.1299{]},\\\quad\,\, $\sigma$={[}0.01, 0.01{]}\end{tabular}     & $1.56^{+0.31}_{-0.53}$ \\[8pt]
Gaussian priors      & \begin{tabular}[c]{@{}l@{}}Mean={[}0.5343,0.1299{]},\\\quad\,\, $\sigma$={[}0.0012,0.0027{]}\end{tabular} & $1.59^{+0.18}_{-0.17}$\\
\hline
\end{tabular}
\end{table}

\noindent \textit{Other noise sources}: Our simulations have considered the ideal situation where only photon (white) noise is present thereby allowing easy scaling of the noise with the number of observations/transits. However in practice, other sources of noise \citep{pont} will impact the estimates we have given above and act to increase the number of transits required to detect deformation. These other noise sources can be from instrumental effects (e.g. satellite jitter and thermal instability) and also from astrophysical sources such as stellar activity (occulted or unocculted active regions \citet{oshagh2013}), stellar oscillations and granulation \citep{chiavassa}. These effects always have to be mitigated in transit analysis \citep{oshagh18,barros14} but will still impact the detectability of shape deformation. Recent development in gaussian process analysis also provides a method for tackling astrophysical noise (e.g. \citealt{dfm17,serrano}).

\section{Conclusion}

Short period planets, especially within 2 Roche radii from the host star, suffer from extreme tidal forces causing their shapes to depart from sphericity in a way that is difficult to detect in transit observations. With the increasing observational precision of near-future instruments, detecting deformation becomes more feasible as the planet shape will have a higher impact on the observed transit light curves. We have demonstrated detectability of deformation for WASP-103b and WASP-121b (which have the highest deformation signatures as seen in Sect. \ref{2.4} and regarded as some of the most deformed planets \citep{delrez}) by employing a formulation from literature in a way that allows an observational estimate of the planet's fluid Love number to be obtained. Because the Love number tells us how a planet deforms in response to perturbing potentials, we used it as a measure of deformation in the planet. Detecting and measuring planet deformation provides more accurate estimations of the radius and density of these planets as opposed to the estimates derived from spherical models or corrections calculated from only expectation of deformation. Additionally, measuring the Love number gives us an information about the interior structure of the planet. We showed that the instrumental precision needed to be attained to detect tidal deformation is $\leq$ 50\,ppm which can be attained by CHEOPS with about 300 transits for WASP-103b and 40 transits for WASP-121b. HST can also attain this precision for WASP-103b in $\sim$40 transit observations. Fewer transit observations will be required if such short-period planets are found transiting very bright stars. Additionally, the precision expected from JWST will present the best opportunity to detect tidal deformation since only one transit of a suitable planet will be required.\\

The chances of detecting deformation is increased for planets with inclinations of $90^{\mathrm{o}}$ and also when the observations are taken with temporal resolution of $\sim$1\,min. However detection can be severely hampered by improper modeling of the limb darkening which, in some cases, can cause the signature of deformation to be subdued leading us to infer sphericity from the observations. Using the quadratic limb darkening law, LDC errors smaller than 0.01 is required in order to confirm planet deformation. Proper treatment of noise sources will also be pertinent in order to identify the signature of shape deformation.  \\

\begin{acknowledgements}
This work was supported by Fundação para a Ciência e a Tecnologia (FCT, Portugal) through national funds and by FEDER through COMPETE2020 by these grants UID/FIS/04434/2013 \& UID/MAT/04106/2013 and POCI-01-0145-FEDER-007672 \& PTDC/FIS-AST/1526/2014 \& POCI-01-0145-FEDER-016886 \& POCI-01-0145-FEDER-022217 \& POCI-01-0145-FEDER-029932 \& POCI-01-0145-FEDER-028953 \&l POCI-01-0145-FEDER-032113. We also acknowledge support from FCT (Portugal) and POPH/FSE (EC) through fellowship PD/BD/128119/2016. NCS and SCCB also acknowledge support from FCT through Investigador FCT contracts IF/00169/2012/CP0150/CT0002 and IF/01312/2014/CP1215/CT0004 respectively. BA acknowledges support from FCT through the FCT PhD programme PD/BD/135226/2017. BA also thanks the LSSTC Data Science Fellowship Program, which is funded by LSSTC, NSF grant \#1829740, and also the Brinson and the Moore Foundations, his time as a fellow has benefitted this work. \\

\end{acknowledgements}

\bibliographystyle{aa} 
\bibliography{references} 

\onecolumn
\begin{appendix}
\section{Additional figures}

\begin{figure*}[!ht]
\centering
\begin{subfigure}{.53\textwidth}
  \centering
 \includegraphics[width=.99\linewidth]{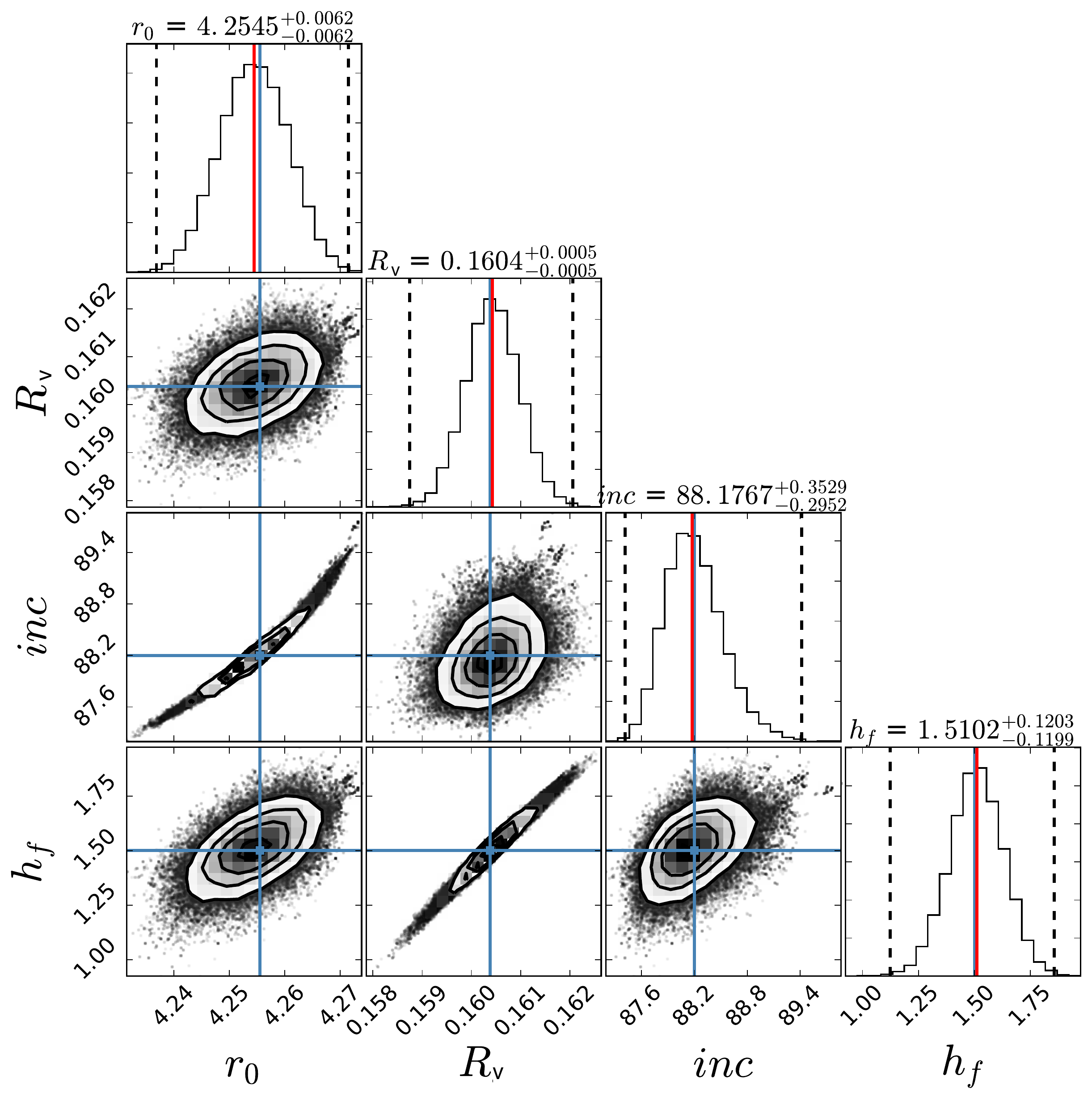}
\end{subfigure}%
\begin{subfigure}{.43\textwidth}
  \centering
 \includegraphics[width=.99\linewidth]{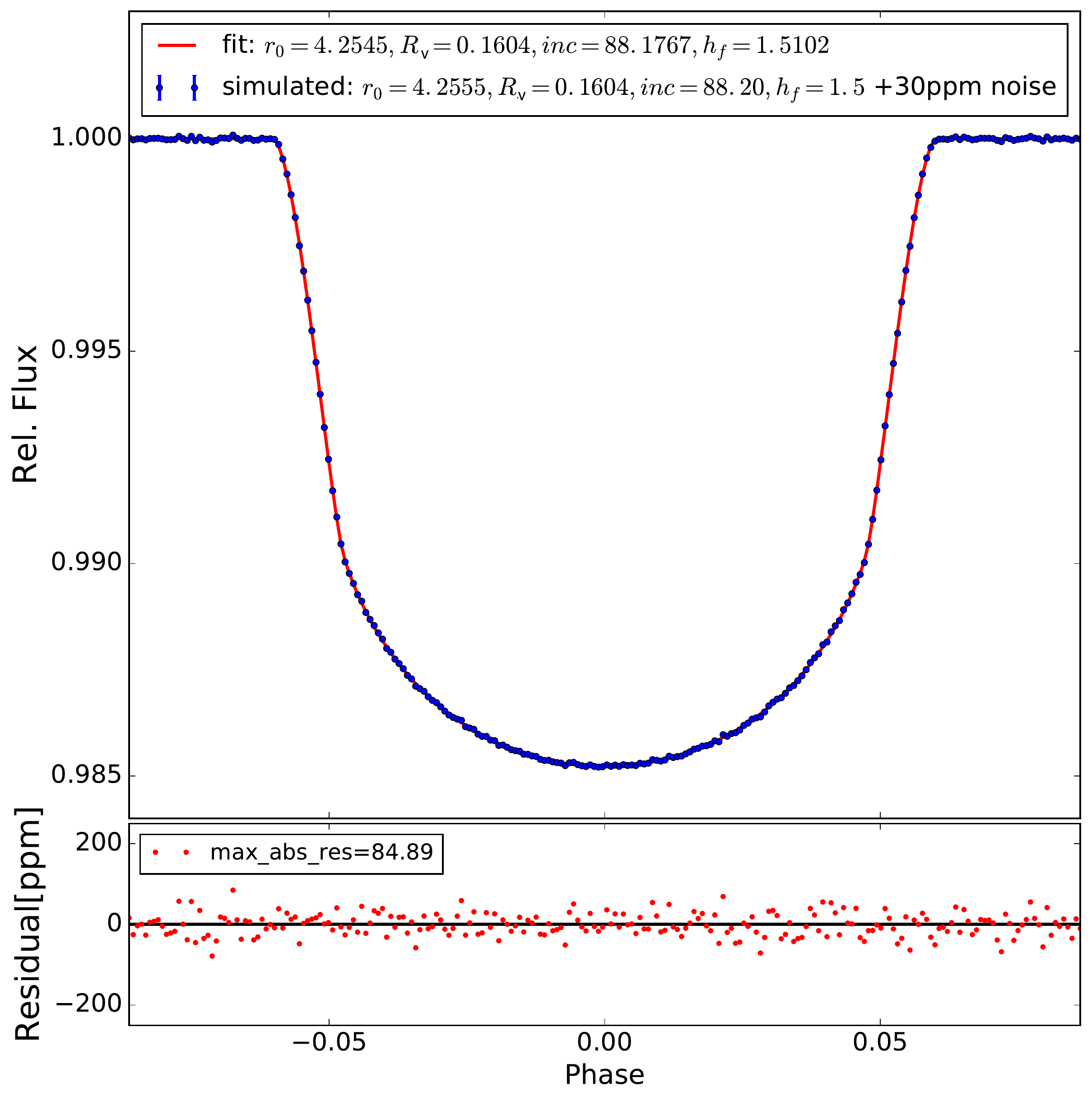}
\end{subfigure}
\caption{\textit{Left}: Posterior distributions for parameters of simulated deformed WASP103b with 30\,ppm noise added. The values quoted on the diagonal histograms indicate the median of each parameter's marginalized distribution (red lines) with the surrounding 68\% credible interval ($\pm1\sigma$). The dashed vertical lines indicate the $\pm 3\sigma$ limits calculated as the $0.15^{th}$ and $99.87^{th}$ percentiles. Blue lines indicate the true simulated values. \textit{Right}: Fit of simulated light curve of  ellipsoidal WASP-103b with parameters retrieved from the sampling.}
\label{corn}
\end{figure*}

\begin{figure*}[!ht]
\centering
\begin{subfigure}{.53\textwidth}
  \centering
 \includegraphics[width=.99\linewidth]{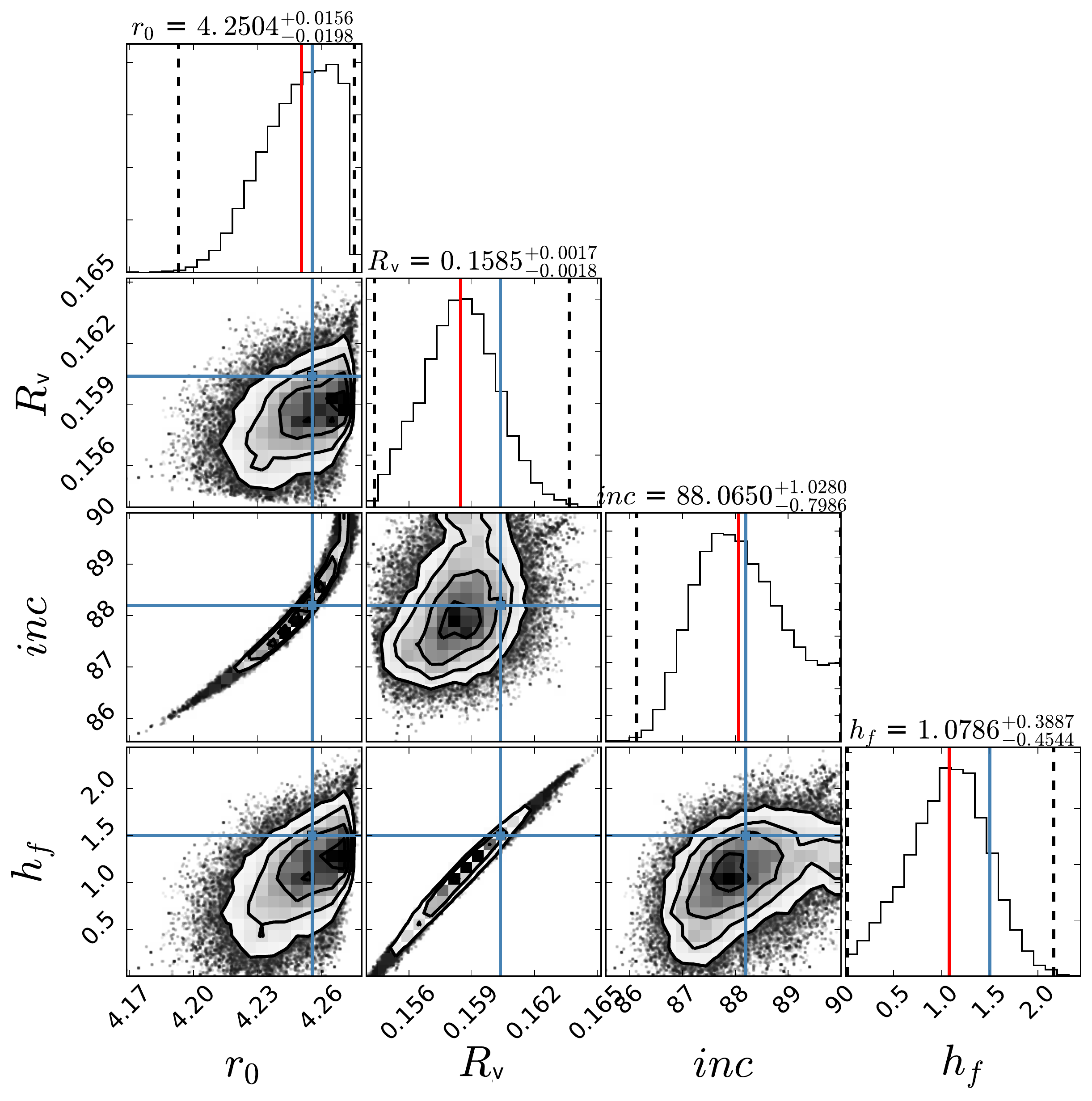}
\end{subfigure}%
\begin{subfigure}{.43\textwidth}
  \centering
  \includegraphics[width=.99\linewidth]{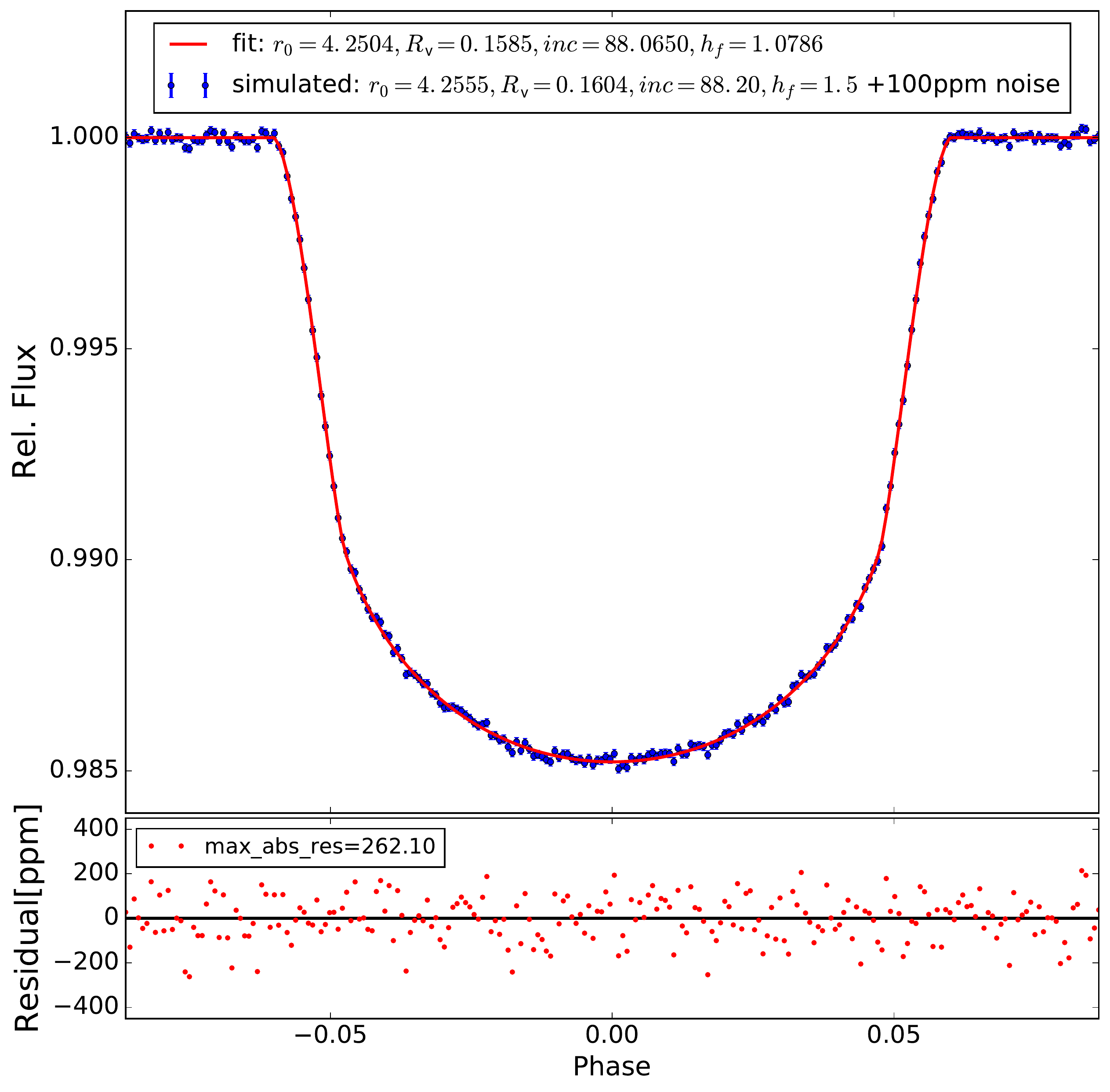}
\end{subfigure}

\caption{Same as Fig. \ref{corn} but with 100\,ppm noise added to the simulated observation.}
\label{corn100}
\end{figure*}

\begin{figure*}[]
\centering
\begin{subfigure}{.50\textwidth}
  \centering
  \includegraphics[width=1\linewidth]{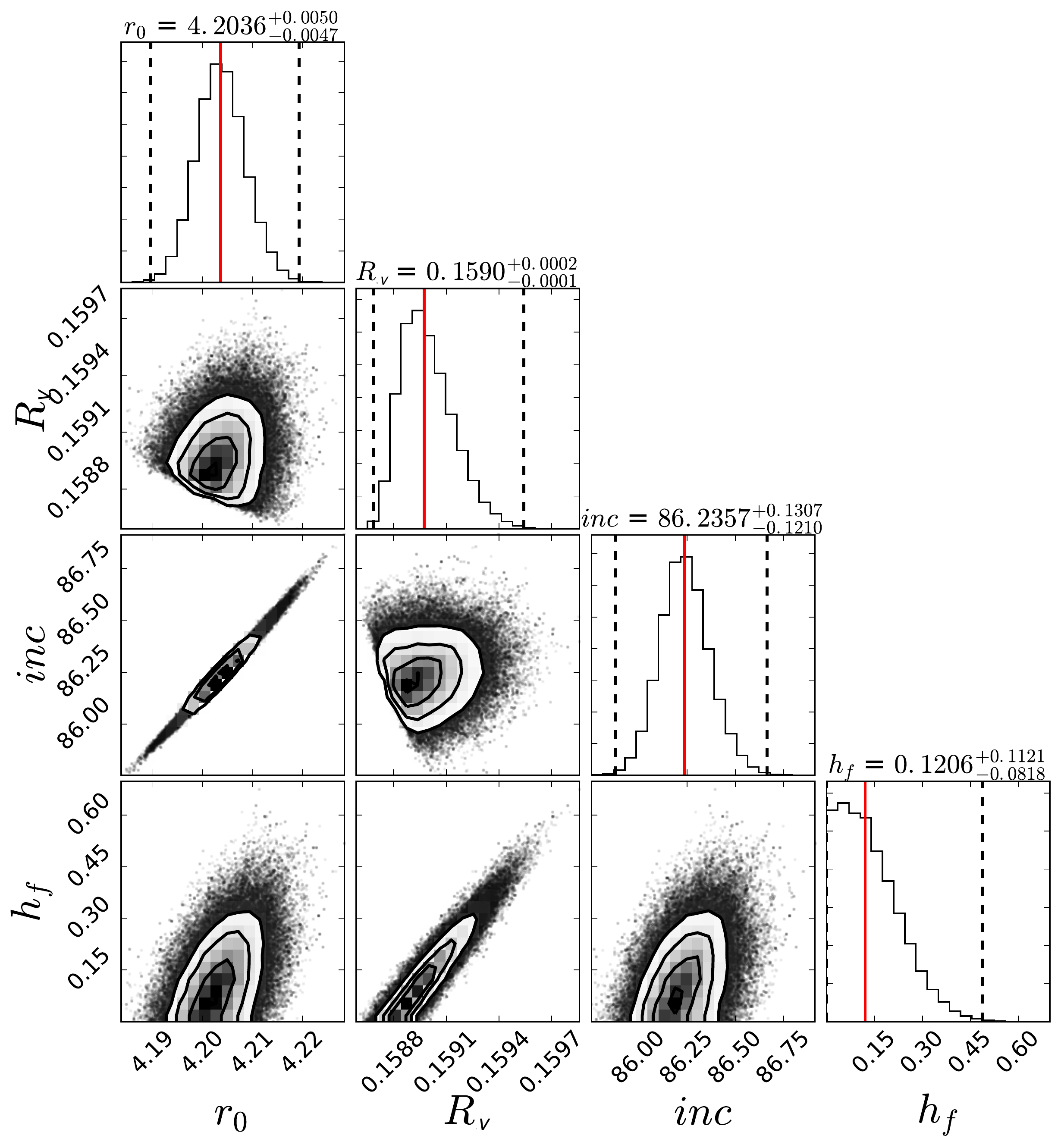}
\end{subfigure}%
\begin{subfigure}{.50\textwidth}
  \centering
  \includegraphics[width=1\linewidth]{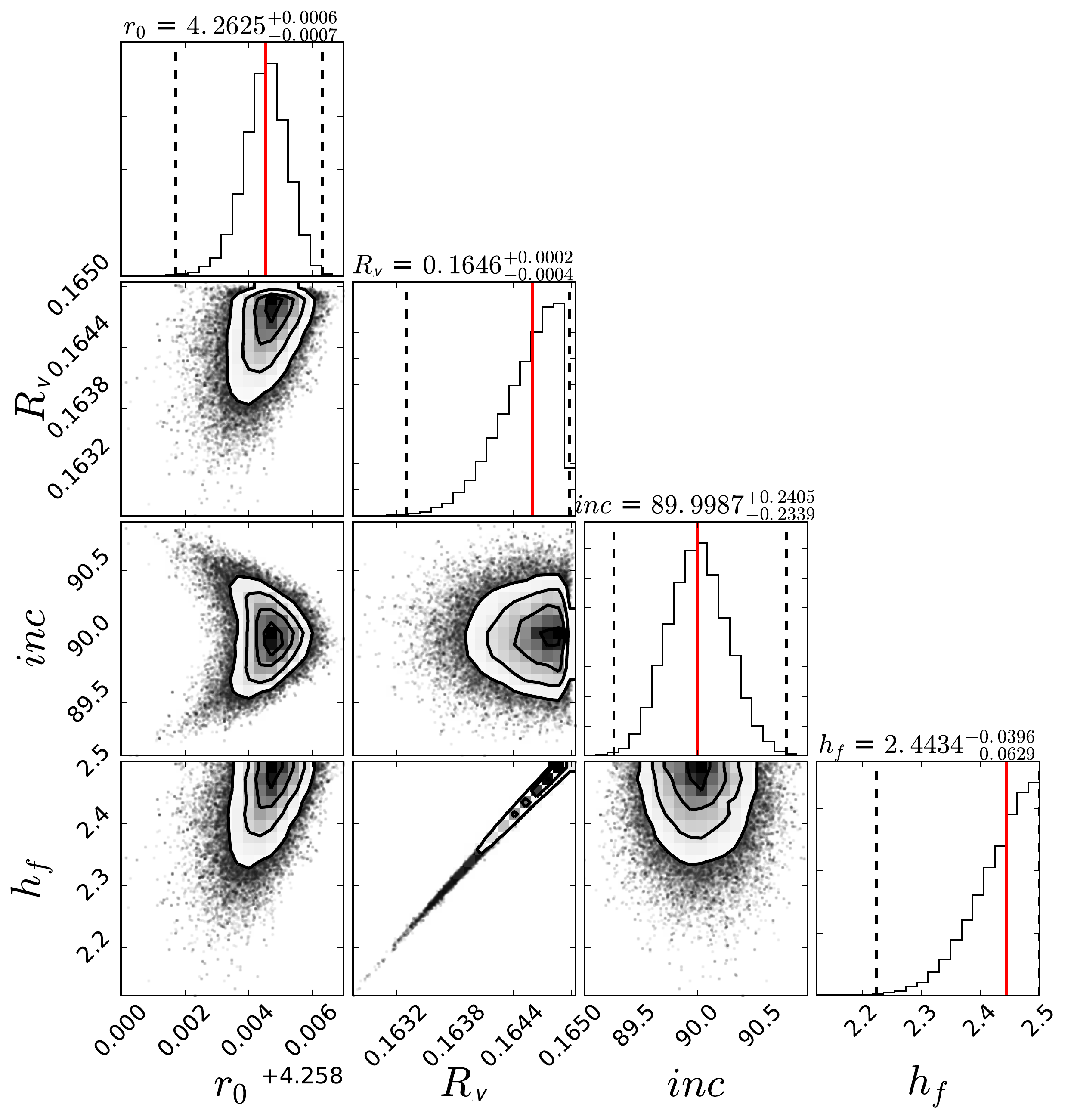}
\end{subfigure}
\caption{\textit{Left}: Same as Fig. \ref{corn} but with LDCs fixed at wrong values 0.01 smaller than the true values \textit{Right}: with LDCs fixed at values 0.015 higher than the true values. }
\label{LDCs}
\end{figure*}

\begin{figure*}[]
\centering
\begin{subfigure}{.51\textwidth}
  \centering
  \includegraphics[width=.99\linewidth]{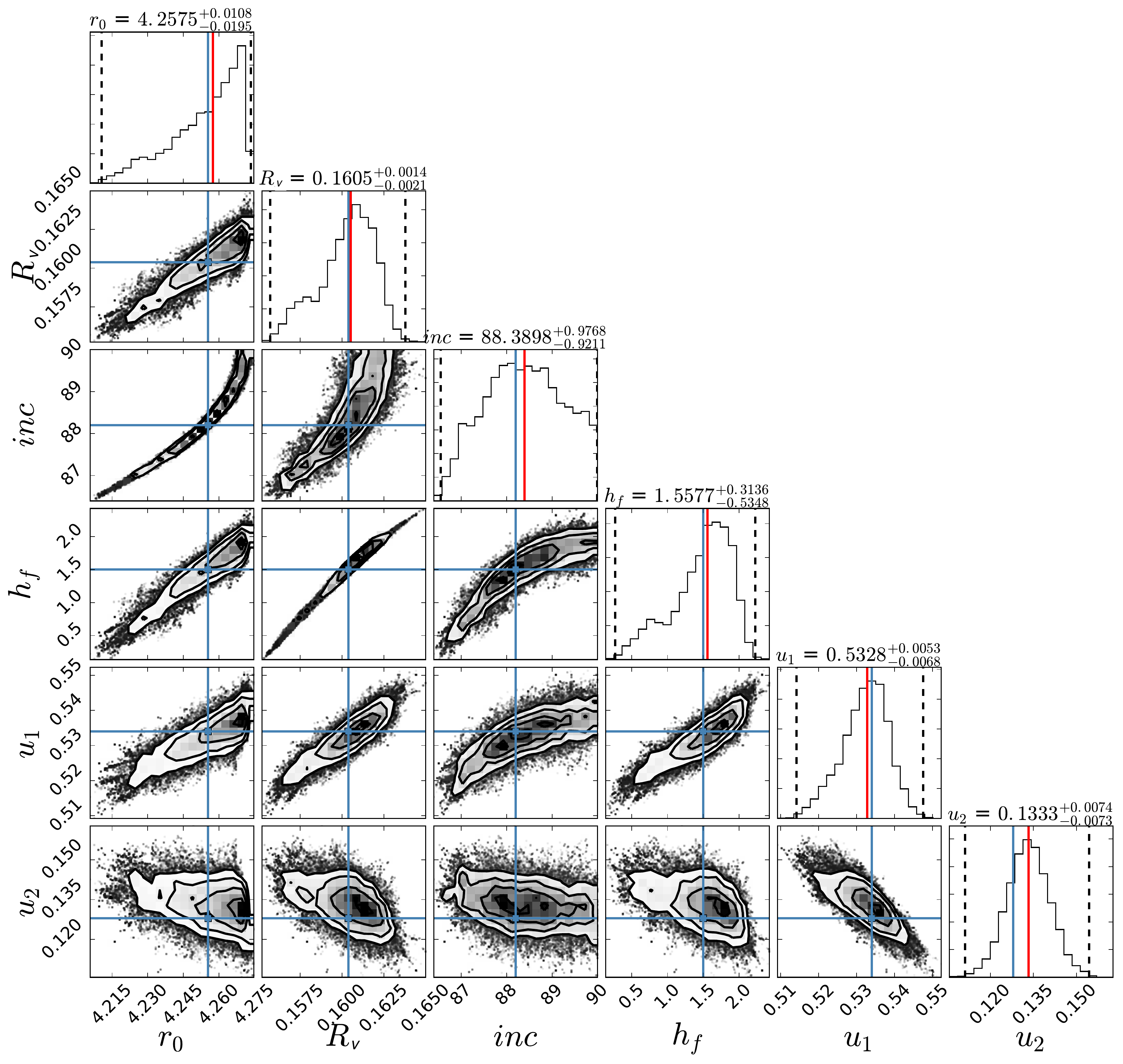}
\end{subfigure}%
\begin{subfigure}{.51\textwidth}
  \centering
  \includegraphics[width=.99\linewidth]{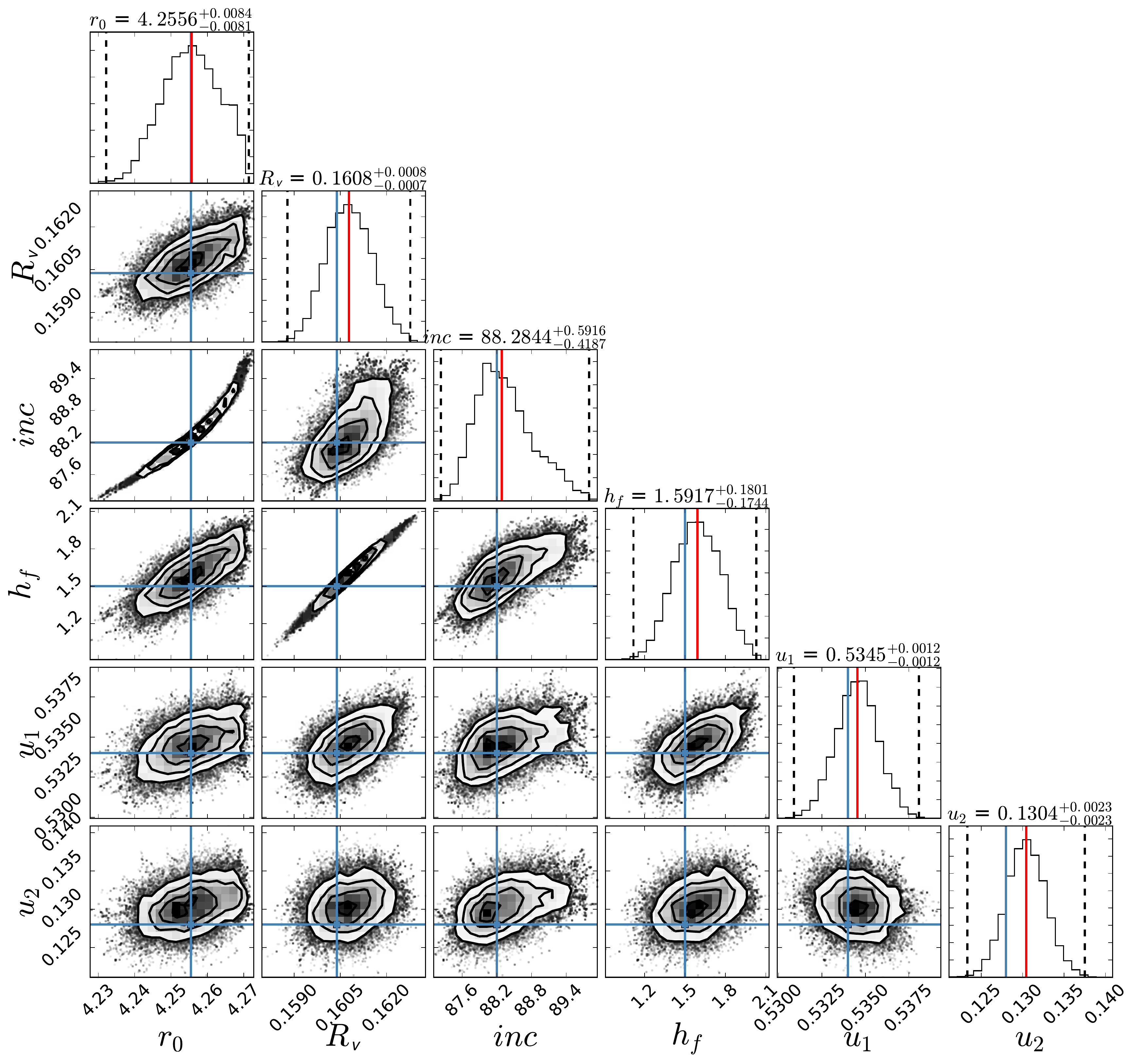}
\end{subfigure}
\caption{\textit{Left}: Posterior distributions of parameters when gaussian prior with $\sigma=0.01$ is used on the LDCs ($u_{1},u_{2}$) \textit{Right}: when tighter gaussian priors from \textit{ldtk} errors are used for the LDCs.}
\label{LDC_priors}
\end{figure*} 

\end{appendix}

\end{document}